\documentclass{bmcart}

\usepackage[utf8]{inputenc} 

\usepackage{bbm}
\usepackage{amsmath}
\usepackage{amssymb}
\usepackage[normalem]{ulem}
\usepackage{multirow}
\usepackage{graphicx}
\usepackage{hyperref}
 
\newcommand\numberOfTweets{15 million}
\newcommand\uniqueUsers{723,000}
\newcommand\numberOfGeoTweets{72,000}

\newcommand\numberOfGeoTweetsMexico{65,000}
\newcommand\numberOfGeoUsersMexico{8,000}

\definecolor{ochre}{rgb}{0.8, 0.47, 0.13}

\startlocaldefs
\endlocaldefs

\begin{document}

\begin{frontmatter}

\begin{fmbox}
\dochead{Research}



\title{Design and analysis of tweet-based election models for the 2021 Mexican legislative election}


\author[
   addressref={aff1,aff2,aff3},                   
   corref={aff1},                       
   email={avigna@mpa-garching.mpg.de}   
]{\inits{A}\fnm{Alejandro} \snm{Vigna-G\'omez}}
\author[
   addressref={aff2,aff4}
]{\inits{JM}\fnm{Javier} \snm{Murillo}}
\author[
   addressref={aff5,aff4}
]{\inits{MR}\fnm{Manelik} \snm{Ramirez}}
\author[
   addressref={aff4}
]{\inits{AB}\fnm{Alberto} \snm{Borbolla}}
\author[
   addressref={aff6,aff4}
]{\inits{IM}\fnm{Ian} \snm{M\'arquez}}
\author[
   addressref={aff7}
]{\inits{PKR}\fnm{Prasun K.} \snm{Ray}}


\address[id=aff1]{
  \orgname{Niels Bohr International Academy, Niels Bohr Institute}, 
  \street{Blegdamsvej 17},                     %
  \postcode{DK-2100}                                
  \city{Copenhagen},                              
  \cny{Denmark}                                    
}

\address[id=aff2]{%
  \orgname{The Aspen Institute M\'exico},
  \street{Av. Cd. Universitaria 298, Jardines del Pedregal,Álvaro Obregón},
  \postcode{01900}
  \city{Mexico City},
  \cny{Mexico}
}

\address[id=aff3]{
  \orgname{Max-Planck-Institut f\"ur Astrophysik}, 
  \street{Karl-Schwarzschild-Str. 1},                     %
  \postcode{D-85748}                                
  \city{Garching},                              
  \cny{Germany}                                    
}

\address[id=aff4]{%
  \orgname{Ciencia de Datos \& Tecnolog\'ia, Metrics, Cd. Sat\'elite, Naucalpan de Ju\'arez},
  \street{Juan Escutia 7},
  \postcode{53100}
  \city{Estado de M\'exico},
  \cny{Mexico}
}
\address[id=aff5]{%
  \orgname{Facultad de Ciencias, Universidad Nacional Aut\'onoma de M\'exico},
  \street{Investigación Científica, Ciudad Universitaria, Coyoacan},
  \postcode{04510}
  \city{Mexico City},
  \cny{Mexico}
}
\address[id=aff6]{%
  \orgname{Facultad de Negocios, Universidad La Salle M\'exico},
  \street{Benjam\'in Franklin 45 Col. Condesa, Del. Cuauht\'emoc},
  \postcode{06140}
  \city{Mexico City},
  \cny{Mexico}
}
\address[id=aff7]{%
  \orgname{Department of Mathematics,Imperial College London},
  \city{London},
  \cny{United Kingdom}
}



\end{fmbox}


\begin{abstractbox}

\begin{abstract} 
Modelling and forecasting real-life human behaviour using online social media is an active endeavour of interest in politics, government, academia, and industry.
Since its creation in 2006, Twitter has been proposed as a potential laboratory that could be used to gauge and predict social behaviour.
During the last decade, the user base of Twitter has been growing and becoming more representative of the general population.
Here we analyse this user base in the context of the 2021 Mexican Legislative Election.
To do so, we use a dataset of \numberOfTweets\ election-related tweets in the six months preceding election day. 
We explore different election models that assign political preference to either the ruling parties or the opposition.
We find that models using data with geographical attributes determine the results of the election with better precision and accuracy than conventional polling methods.
These results demonstrate that analysis of public online data can outperform conventional polling methods, and that political analysis and general forecasting would likely benefit from incorporating such data in the immediate future.
Moreover, the same Twitter dataset with geographical attributes is positively correlated with results from official census data on population and internet usage in Mexico.
These findings suggest that we have reached a period in time when online activity, appropriately curated, can provide an accurate representation of offline behaviour.
\end{abstract}




\end{abstractbox}
%

\end{frontmatter}





\section{Introduction}
In 1824, newspapers reported arguably the earliest public opinion polls in the context of the United States presidential election campaign of the same year \cite{tankard1972public}.
Since then, the use of diverse forms of media, such as newspapers, radio, and television, have been systematically used to track and discuss elections all over the world.
The more recent dawn of social media led to the rapid development of the online arena as the preferred venue for large-scale political discussions and campaigning \cite{Tumasjan_Sprenger_Sandner_Welpe_2010,ICWSM101536,bond201261,digrazia2013more,BURNAP2016230}.
The microblogging and social networking service Twitter has become increasingly relevant in political discussions, both from the perspective of engaging citizens \cite{digrazia2013more,BURNAP2016230,bovet2019influence} and campaigning politicians \cite{dimitrova2018social,doi:10.1080/15377857.2016.1228555}.
In the early days of Twitter, imbalanced representation and population biases were a major concern for modelling and forecasting \cite{kohut2012assessing,barbera2015understanding}.
The increase of internet users worldwide \cite{ITU2021,pew2021} with hints of increasing homogenization among Twitter users \cite{Nishida2018Politics,pew2019} highlights the potential of social media as a powerful tool for political and societal analysis \cite{Conover2012,MussiReyero2021}.

Twitter data has been used to study a wide range of topics, including troll activity \cite{alizadeh2020content}, cognitive reflection \cite{2021NatCo..12..921M}, digital trace data to study migration and migrants \cite{Armstrong2021}, expressed sentiment alterations during the COVID-19 pandemic \cite{Jing2021,wang2022global}, misinformation spread during earthquakes \cite{flores2021fighting}, and spatial analysis of gunshots reports \cite{garcia2021spatial}.
In the context of politics, there has been focus on the influence of fake news \cite{bovet2019influence,grinberg2019fake}, campaigning \cite{dimitrova2018social,Bright2020elections,doi:10.1080/15377857.2016.1228555}, echo chambers \cite{doi:10.1177/0956797615594620}, polling \cite{bovet2018validation}, and elections \cite{khan2021election}.
For online polling and election predictions the preferred methods of analysis are based either on sentiment \cite{BURNAP2016230}, volume \cite{digrazia2013more}, or social networks \cite{bovet2018validation}, and hybrid methods are the exception rather than the rule \cite{khan2021election}.
These analyses have led to a wide variety of results, including those where Twitter opinion predicts the results from aggregated polls \cite{bovet2018validation} or outperforms polls \cite{zhenkun2021polls}, but also those where forecasting failed to predict the broad outcome of an election \cite{BURNAP2016230}
(see the recent surveys by \cite{khan2021election,chauhan2021emergence,9383789,santos2021survey} for a thorough review of the use of social media data for election predictions).

In this paper we gauge \textit{representativeness} of online activity with respect to real-life behaviour.
We compare Twitter data with conventional aggregated polls \cite{oraculus2021} and official results \cite{INE2021} of the 2021 Mexican legislative election (Sect. 2.1).
In Sect. 3 we present our methodology, particularly data mining (Sect. 3.1), allegiance determination (Sect. 3.2), and election models (Sect. 3.3).
In Sect. 4 we present our results, which are discussed in Sect. 5.
Finally, Sect. 6 presents the concluding remarks of this paper.

\section{Background}
The 2021 Mexican legislative election was a federal election that took place in Mexico on June 6, 2021 (Election Day).
In this election, 500 seats were elected and allocated in the Chamber of Deputies.
This election takes place every three years and is organised and overseen by the National Electoral Institute (INE, for its acronym in Spanish).
The INE is an autonomous, public agency in charge of organising and reporting the results of federal elections.
Additionally, the INE is in charge of approving coalitions made among parties.
In Mexico, coalitions among parties have become increasingly common in the last decades.

For the 2021 Mexican legislative election, there were two major coalitions, of three parties each, and four additional individual parties.
The coalition ``Juntos Hacemos Historia" was comprised of the parties MORENA, PT, and PVEM.
At the time of the election, MORENA was the party which held the majority of seats in the Chamber of Deputies.
Moreover, the president of Mexico, elected on 1 December 2018, has been affiliated to MORENA since the gestation of the party in 2011.
Therefore, here and hereafter, we refer to the coalition Juntos Hacemos Historia as the \textit{ruling parties}.
The coalition ``Va por M\'exico" was comprised by the strong alliance of (historically opposing) parties PAN, PRI, and PRD.
In order to analyse this multi-party election, we opt for a bipartisan model where the coalition ``Va por M\'exico" and the remaining, smaller parties (MC, PES, FxM, and RSP) are agglomerated as the \textit{opposition}.

\section{Methods}
We queried Twitter via the Twitter API for Academic Research to retrieve election-related tweets (Sect. 3.1.1).
Our goal is to compare Twitter data to both official results and aggregated polls (Sect. 3.1.2).
We process the Twitter data in order to determine the allegiance of each tweet with respect to the ruling parties or the opposition (Sect. 3.2).
We then construct nine different models of the election (Sect. 3.3).

\subsection{Data Mining}
\subsubsection{Twitter}
We retrieved data from Twitter using the Twitter API for Academic Research via the \textit{twarc} Python library \cite{ed_summers_2022_6503180}.
Between November 2021 and February 2022, we queried and retrieved the pertinent tweets posted between December 1st, 2020 and May 31st, 2021 at 05:00:00 UTC, which corresponds to midnight in Mexico City.
This corresponds to the 6 complete months preceding Election Day in June 6, 2021.
We make individual, tailored queries for each of the 10 individual parties in the 2021 Mexican legislative election \cite{INE2021}.
We agglomerate the individual parties in a bipartisan model: the ruling parties vs the opposition.
The ruling parties are comprised of 3 individual parties with the following acronyms: MORENA, PT, and PVEM.
The opposition is comprised of 7 individual parties with the following acronyms: PAN, PRI, PRD, MC, PES, FxM, RSP.
We tried to keep the queries as simple and concise as possible \footnote{The string query for each search done to analyze the data in this manuscript is publicly available via Zenodo \cite{vigna_gomez_alejandro_2022_7877001}.}.
All queries contain at least the name of the party and, when available, the verified Twitter handle.
Most queries also included hashtags; these could be as simple as a reference to the queried party (e.g. \#PAN for PAN) or include hashtags frequently used by the official Twitter accounts (e.g. \#SomosPT as used by PT).
The case of MORENA is a peculiar one, as the acronym (intentionally) means ``brunette" in Spanish.
This query requires some additional filters, determined by doing manual tests on small data and excluding keywords from the query, that retrieve almost exclusively tweets associated to the election.
As a result, this query is \textit{ad hoc} and long  \cite{vigna_gomez_alejandro_2022_7877001}, but within the allowed number of characters.
Finally, during post-processing (i.e. not during the extraction via the Twitter API) all queries were restricted to results in Spanish.
This results in $\approx$\numberOfTweets\ tweets from $\approx$\uniqueUsers\ unique users posted between December 2020 and May 2021.

\subsubsection{Official results and polling aggregates}
We use the official vote count as provided by the INE \cite{INE2021}.
For the polling aggregates, we use publicly-available data as provided by \textit{oraculus}, a website which specializes in political analyses \cite{oraculus2021}.
The data provided by oraculus contains public polls from news outlets and market research agencies.
The data includes a total of 26 polls between December 2020 and May 2021, comprised of both phone and in-person polls.
We use the \textit{effective} data, which does not incorporate undecided nor unresponsive interviewees.
Oraculus uses the effective data from November 2018 to May 2021 to build a Bayesian model.

\subsection{Allegiance determination}
Our modeling analysis relies on determining the allegiance of each individual tweet, including retweets, with respect to the query from where it was retrieved.
Figure 1 presents the allegiance distributions of the ruling parties and the opposition, in the form of violin plots, for different subsets of the data of May 2021 (the month preceding Election Day).
We determine the allegiance in the following way.
First, we create a matrix of tweets that we tokenize using the Tokenizer Package from the Natural Language Toolkit (NLTK) \cite{bird2009NLTK}.
Then we convert the text into a matrix using the CountVectorizer function from the scikit-learn Python library \cite{scikit-learn}.
We proceed to use the multinomial Naive Bayes classifier, from the scikit-learn library, in the tokenized text data to estimate the probability of the tweet to have a positive connotation.
This probability is close to 0(1) when the tweet expresses a negative(positive) connotation.
For all parties we used supervised machine learning training with manually classified tweets.
The manually classified tweets are categorized as either positive (p) or negative (n).
We use a tweet dataset pertinent to the election to train our model by manually categorizing a fraction of the tweets (85\%) and then use the rest of the data (15\%) to test our trained model.
Each party was classified using an individual training set with $\sim 100-1000$ messages.
Table 1 shows the summary of our training models.
We have also tested the effectiveness of a pre-trained transformer model \cite{canete2020spanish}, however we have found that our Naive Bayes Classifier is better at correctly classifying tweets that discuss simultaneously the ruling parties and the opposition.
The output of our algorithm retrieves a matrix that, for each tweet, indicates the following attributes: tweet ID, user ID, region, country, party-of-interest (e.g., PAN), estimated allegiance ($0.0\leq \mathcal{A} \leq 1.0$), date, and coalition.
This matrix is the main output we use to perform the analyses in this manuscript.

\subsection{Election models}
In order to compare our Twitter data with official results and estimates from polls, we explore different election models from the literature.
These models capture different nuances in their assessment of when and how can a tweet be representative of a vote.
Arguably, the simplest approach is to use a volumetric tweet-based model (VT) where each tweet that mentions a party is assigned a \emph{vote} for that party's coalition \cite{digrazia2013more}. 
This model has obvious shortcomings; for example, a single enthusiastic user can easily generate a large number of tweets which are reflected as votes for a party. 
This problem is alleviated by using a volumetric user-based \cite{gaurav2013leveraging} model (VU) where each account that mentions parties in one coalition more often than those in the other coalition is counted as a \emph{voter} for that coalition. 
However, both volumetric approaches ignore tweet sentiment. 
A tweet mentioning a party could be expressing a negative sentiment; Figure 1 shows that a large proportion of the tweets in our dataset fall in this category. 
Consequently, we introduce a suite of models that incorporate results from sentiment analysis into their vote-share predictions \cite{BURNAP2016230}.
First, an allegiance score, $A^{(y)}$, between $0$ and $1$ is computed and assigned to each tweet that mentions a party in coalition $y$. Here $A^{(y)}=0$, $A^{(y)}=0.5$, and $A^{(y)}=1$ correspond to negative, neutral, and positive attitudes towards $y$, respectively.  
We then compute vote-share predictions based on tweets (AT) and users (AU). 
The former approach simply sums the allegiance scores for each coalition and the ratio of these scores is used to estimate their vote shares. 
The user-based approach computes an average allegiance for each user and for each coalition. 
These averages are then totalled and the ratios of these totals is used to estimate vote shares.

Both the volumetric and sentiment-based approaches fail to consider if Twitter users are actually representative of the general electorate. 
We propose two modifications to the sentiment approach to address this issue. 
The first alternative modifies the previously described models by only considering the subset of tweets that carry geolocation data, referred to as \textit{geodata} hereafter. 
This allows us to compare the geographical distribution of people with our tweets and thus quantify one aspect of representativeness. 
In our results, models labeled with a \emph{G} prefix use this subset while those labeled with a \emph{C} prefix use the complete dataset.
The second \emph{alternative} model assumes that the subset of users posting predominantly positive tweets are more representative of the electorate; we will find this alternative model produces a good estimate of the election results. 

The alternative election model does not rely on geodata and focuses on the online positive allegiance, per month, of unique users to a party.
Vote share estimates in this model are constructed as follows.
First, we label the tweets depending on whether they reference the ruling parties (y=0) or the opposition (y=1); tweets which reference both parties simultaneously receive both labels.
For each label we average the allegiance of tweets per unique users in the form $\overline{\mathcal{A}}^{(y)}_i := \sum_{n=1}^{N^{(y)}}\mathcal{A}^{(y)}_{i,n}/N^{(y)}$, where $i$ is the index of a unique user, and $N_i^{(y)}$ is the number of tweets posted by user $i$ with label $y$. 
We then focus on users that express a positive connotation, exclusively, to either the ruling parties or the opposition; i.e., if one user expresses positive attitude about both the ruling parties and the opposition they are excluded from the subsequent analysis.
A positive connotation is defined by setting lower ($x_{\rm{low}}$) and upper ($x_{\rm{upp}}$) bounds for user allegiances and filtering them by considering only the users within those limits.
Therefore if $x_{\rm{low}} \leq \overline{\mathcal{A}}^{0}_i \leq x_{\rm{upp}}$, and $\overline{A}_i^{(1)}<x_{low}$, the model determines that user $i$ will vote in favour of the ruling parties.
The default model uses $x_{\rm{low}}=0.6$ and $x_{\rm{upp}}=1$, however we do present results where other values have been used (in the ranges, $0.1 \leq x_{\rm{low}} \leq 0.7$ and $0.7 \leq x_{\rm{upp}} \leq 1$).

\section{Results}
\subsection{Election model comparison}
Figure 2 shows the estimated vote share for the ruling parties for all of our models during May 2021.
For each of our models, we do 1000 random samplings with replacement on the monthly data, estimate the mean value for each sample, and present the vote-share distribution as box plots defined by the median and the lower and upper quartiles. 
Additionally, we present the reported official results from the election (44.37\% for the ruling parties) and from aggregated polls (49\% for the ruling parties).
Throughout the six months preceding the election, the vote-share estimate fluctuates (Figure 3).
Some models differ significantly, with preference for the ruling party being as low as 34.0\% in December 2020 according to the alternative model and as high as 86.6\% in March 2021 for the CVT model.
However, some features are shared among all models.
The political preference for ruling parties increases until March or April 2021 and then drops and converges.
For polling aggregates, this drop happens earlier, in February 2021, resulting in convergence since March 2021.
This fluctuating evolution of political preference could be, at least in part, due to the large amount of swing voters, which are the voters who decide on how to vote late on the election \cite{hargittai2018biases}.
Models which use geodata and the alternative positive-allegiance analysis differ from the official results by $\lesssim$3.6 percentage points.
They are more accurate than conventional polling methods, which differ to official results by $\approx$4.6 percentage points, and significantly more accurate than models based on the complete data.
The precision of all models ($<5.5$ percentage points), i.e., the width of the distribution or the size of the boxplot, is better than that of aggregated polls ($\approx 5.8$ percentage points).
Figure 4 shows the monthly evolution of the number of tweets depending on the model.
Our models suggest that $\lesssim$100,000 users (63,500 for the alternative model) are representative of voting intention in Mexico, and even as little as $\lesssim$10,000 users (5,200 in the GVU and GAU models) from a Twitter dataset with geodata can be used to model elections.
For comparison, in the voting booth the total number of votes was $\approx$49 million out of $\approx$93 million registered voters \cite{INE2021}.
Examining the model vote-share predictions, we observe that the tweet-based (T) predictions are closer to the actual vote share than the user-based (U) predictions. 
Similarly, the allegiance-based models (A) tend to outperform the volume-based models (V). 
Particularly, the alternative allegiance-based model performs much better than most other models based on the complete data set.
Most importantly, the differences between these models becomes much smaller and the predictions tend to become more accurate when geodata is used.

The alternative positive-allegiance model proposed here outperforms aggregated polls and most election models, including some that rely exclusively on geodata (Figure 2).
For this model, we perform a detailed analysis of the uncertainties in accuracy, precision, and volume for the data of May 2021 (Figure 5).
For each pair of limits, $\{ x_{\rm{low}}, x_{\rm{upp}} \}$, we obtain a sub-model of the election. 
The accuracy of our sub-models improve when increasing both limits, which implies that the more we focus on users discussing the election positively, the better we are able to match the results of the election.
The uncertainties of the alternative model is $\lesssim 5$ percentage points for all sub-models.
The volume of unique users increases proportionally with the width set by the limits, spanning from $\approx$12,000 to $\approx$169,000 unique users.
For the alternative model, with $\overline{\mathcal{A}}\geq 0.6$, 63,500 unique users result in an accurate and precise $0.4\pm 2.3$ percentage points difference with respect to the official results of the election.

\subsection{Geo-analysis}
Our results suggest that tweets with geodata may have been representative of voter intention during the 2021 Mexican legislative election (Figure 2).
We proceed to perform a quantitative geographical analysis of our geodata and compare it to results from the official census to assess the representativeness of our dataset.
Figure 6 shows a barplot of the geographical analysis of the population of Mexico and our Twitter data.
We compare the population \cite{inegi2020}, number of internet users \cite{ENDUTIH2020}, and location of Twitter users for each of the 31 states and capital city (Mexico City) comprising Mexico.
In our data, \numberOfGeoTweets\ tweets out of a total of \numberOfTweets\ have geodata, i.e. $\approx$0.5\%.
The number of tweets with geodata exclusively from Mexico is $\approx$\numberOfGeoTweetsMexico, which corresponds to $\approx$\numberOfGeoUsersMexico\ unique users; we focus on these unique users for our estimates.
The distributions of both internet and Twitter users across the 31 states shows good agreement with the state populations.
For Mexico City, the percentage of inhabitants (7.3\%) and internet users (8.6\%) is quite close, while the percentage of Twitter users from our geodata is over represented (20.4\%).
In order to quantify this level of agreement, we calculate the Pearson’s correlation coefficient (r) between the different populations.
The value between the percentage of the population and percentage of internet users is $r=0.98$, and decreases to $r=0.67$ when comparing to the Twitter data (the Pearson coefficient of the population of internet and Twitter users is $r=0.73$).
This value increases to $r=0.96$ when we combine the data from Mexico City (MX), Hidalgo (HG) and the State of Mexico (MC), which encompasses the conurbation around Mexico City known as Greater Mexico City.
This confirms that the population and internet users are highly positively correlated, and the Twitter population from our data is positively correlated.
We have also examined the \emph{residuals} for each region by subtracting the percentage of Twitter or internet users from the percentage of the overall population.
These residuals show that in most, but not all, of the states the data is under-represented.
This is an obvious feature for the Twitter data in which Mexico City is over-represented, as the other states need to be underrepresented for the total to add up to 100\%.
Outside Greater Mexico City, the internet and Twitter residuals are within 1.6 and 3.0 percentage points, respectively.

We perform an additional analysis on the data to explore representativenes in the geodata.
Figure 7 explores each model with geodata to reproduce the actual population distribution of each state.
To do so, we use a subset of the model data to sample 1000 users which follow the real distribution.
We repeat this process 1000 times (bootstrapping with replacement) to get a mean of the vote share for the ruling parties and create distributions to explore the uncertainties.
These are presented as boxplots, similar to the presentation of our main results (Figure 2).
The original data and the geo-corrected distributions are in very good agreement.
Uncertainties in models with the geo-corrected distributions ($\sim 1,000$) is larger than in the original data ($\gtrsim 1,000-10,000$) given that they are subset of it (Figure 4).

\section{Discussion}
Here we have shown that, for the 2021 Mexican legislative elections, models based on a positive-allegiance analysis or using geodata perform very well, while models using the complete data perform poorly. 
The positive-allegiance model is simple to implement and is also more intuitive: if a user, on average, expresses themselves positively towards a party, they are likely supporting that party and would vote for them if they had the chance. 
However, the discrepancy between the models using complete or geodata is more difficult to explain. 
For both data, the bulk of the allegiances lies close to zero (Figure 1), implying that most tweets are expressing something negative about the pertinent party. 
However, the same models using complete or geodata differ by 10 or 15 percentage points (Figure 2). 
These results could suggest that geodata i) is more representative to reality, ii) provides users more trustworthy opinions, or iii) filters out bots and trolls. 
We find it difficult to point out exactly the reason why models with geodata outperform those which use the complete data, and therefore refrain from making any conclusive remarks.

\subsection{Polling, Twitter, and open-source intelligence (OSINT)}
Concerns about the accuracy of conventional polling have been appearing in major news outlets in recent years \cite{NYTimes2018,NYTimes2022}.
For example, phone-based polling now faces challenges that were either weaker or absent in previous decades. 
Our study used archival Twitter data to gauge representativeness of online activity with real-life decisions.
The Twitter data was used to determine political preference and regional volumetric activity, which we then compared to the official election results and census.
Our main finding is that bipartisan models of the Mexican election using geodata are more accurate and more precise than conventional polling methods, including Bayesian modeling of aggregated polls \cite{oraculus2021}.
These findings build on previous evidence that Twitter data, used in the context of elections, not only matches national polling aggregates, but precedes their results by days \cite{bovet2018validation}, highlighting the power of real-time communication.
The advantages of online polling are well known: it is quicker, cheaper, and reaches a large number of people \cite{hargittai2018biases,bovet2018validation}.
Our approach is different from polling, whereas instead of directly asking someone their opinion we build a contextual query of the topic of interest and retrieve the pertinent data; this is more similar to OSINT methods.
The main criticism towards modeling and forecasting using online resources, in this case Twitter, is about the known and unknown biases from the data \cite{holbrook2010social,Buskirk2022}.
However, OSINT-like methods can be useful to get around some biases from conventional polling \cite{zhenkun2021polls}, such as social desirability bias
\cite{crowne1960new,fisher1993social}, which is a tendency of survey respondents to answer in a way that they consider socially favourable rather than with their actual opinion.
Other systematic uncertainties are shared between conventional polling and OSINT-like methods, such as over-reporting, which occurs when people respond to a poll or engage in online political discussions but do not end up voting \cite{silver1986overreports}.
Another known problem has been the use of a pure volumetric approach where more tweets are assumed to be directly correlated to more votes \cite{digrazia2013more} and where the quality of predictions has been mixed.
In our study, we have also included models which estimate vote share based on the inferred voting intent of unique accounts, an approach that weighs equally the average opinion of very active accounts and those that are less active.
Finally, the demographics of users from Twitter or other social media is an ongoing topic of debate. Most of the analyses around demographics have been made around Western societies, and particularly in the United States. 
Americans that do not use the internet are, in their majority, 65 years or older, earn less than 30,000 USD per year, and did not go to University \cite{pew2021}; however, the number of Americans that are offline has increased from $\approx50\%$ in 2000 to 7\% in 2021 \cite{pew2021}.
There is evidence that ideological segregation in social-media usage between left and right has been overestimated \cite{doi:10.1177/0956797615594620}, that densely populated areas tend to be underrepresented consistently in non-spatial models \cite{ijgi10050323}, and that cognitive reflection correlates with behaviour \cite{2021NatCo..12..921M}, which could be linked to socio-demographic variables such as age, income, and education.
Analyses on representativeness have been conducted in other countries, even if they are less common.
In Japan it has been pointed out that ``In the early days of the Internet in Japan, there was a temporary “liberal bias” because users were skewed toward the urban and highly educated segments \cite{kobayashi2007socialization}. However, this gap has now disappeared, and conservative parties such as the incumbent Liberal Democratic Party have adapted better to the online environment \cite{Nishida2018Politics}." \cite{yoshida2021japanese}.
In Mexico, representativeness has been explored in the context of Slacktivist \cite{howard2016social} and data voids \cite{flores2022datavoidant}, leaving room for future analyses on a more general representativeness.

\subsection{Bots}
Nowadays, the presence and activity of automated accounts (bots) is one of the main topics of debate regarding Twitter \cite{Woolley_2016}.
In the past, Twitter has reported that a few percent ($\approx$5) of all accounts are bots.
However, different academic studies suggest that the percentage of bots may be as high as $\approx$15 \cite{varol2017online,RODRIGUEZRUIZ2020101715}.
The unequivocal classification of bot-like behaviour and bot accounts is non-trivial.
Moreover, not all bot accounts are malicious, simultaneously active, or participating in political propaganda.
Studies have explored that bots that do participate in political propaganda \cite{forelle2015political} tend to be spread across the whole political spectrum \cite{Bruno2022} and play a role in amplifying an exchange of content rather than creating a horde of partisan followers \cite{caldarelli2020role}.
There is tentative evidence that verified accounts play a stronger role than bots during contentious political events \cite{gonzalez2021bots}.
We did not incorporate detection and filtering of bots in our analysis.
Focusing on unique users and neglecting extremely positive or negative allegiances is useful to marginalise bots; both of these approaches are incorporated in our alternative model.

\subsection{Comparison with the literature}
The role of social media in elections has been a topic of interest and investigation for more than a decade now \cite{ICWSM101536,bond201261,karpf2012moveon}.
Most analyses have been performed in the context of the most powerful (and digitally connected) economies of the world, such as the United States \cite{digrazia2013more}, the United Kingdom \cite{BURNAP2016230} and Germany \cite{Tumasjan_Sprenger_Sandner_Welpe_2010}.
One of the seminal studies that explored the role of Twitter in Latin America made predictions, via a volumetric analysis, for elections in Venezuela, Paraguay, and Ecuador \cite{gaurav2013leveraging}.
In Mexico, the connection between politics and social media has been investigated in the context of militias \cite{savage2015participatory}, civic engagement and slacktivism \cite{howard2016social}, and disinformation on political topics in the media \cite{flores2022datavoidant}, but predictions and quantitative analyses of Mexican elections are rarely found in the literature.
Recently, a machine learning approach was used to predict four presidential elections in Latin America in 2018 \cite{brito2023machine}.
This method collected 65,000 Facebook, Twitter, and Instagram posts that were jointly analysed with polls. 
For Argentina, Brazil, and Colombia, the analysis was successful, with predictions similar to the results of the elections.
However, that was not the case for Mexico, where the winning candidate was under predicted by more than 10 percentage points.
The authors were not able to identify the reason for this, but they do suggest that it could be related to a lack of data.
Specifically for Mexico, they collected 9843 posts, 7146 which were from Twitter, spanning a similar timescale to that of our study; however, our (geo)data is (one)three orders of magnitude larger than the data collected for their study.
Additionally, they combine the data from different social media; while they consider this homogenised their sample, we consider that it introduces more uncertainties in the joint analysis and adds more caveats to the interpretation of their results.  

\subsection{Methodological choices and their challenges}
Here we address some of the nuances and challenges related to election modelling with social media data. 
Data mining is arguably the most difficult aspect of our analysis. 
It involves generating queries with keywords, analysing the retrieved data, and then further refining the original query to exclude unrelated tweets. 
In our pipeline, choosing exclusively tweets in Spanish helped us to vastly reduce tweets that had nothing to do with the election.

For our analysis we decided to perform a bipartisan study instead of an independent party study. In a multi-party election that allows for party coalitions, the most straightforward way to perform the analysis is per party or per coalition. 
With tweets, one needs to be careful, as a single tweet can make reference to multiple parties, multiple coalitions, and a mix of parties and coalitions. 
If the analysis is done per party, tweets that exclusively mention the coalition or mention more than one party in the coalition should be weighted or ignored. 
If the analysis is done per coalition, the role of individual parties without coalitions must be assessed. 
A bipartisan analysis allowed us to adequately incorporate the role of coalitions and provided a simple framework for comparing a broad range of election models. 
In the literature, a frequent approach for multi-party election studies is to perform an analysis of all parties or candidates, but focus on presenting the results for the most dominant party or candidate \cite{brito2023machine}. 
Future studies on elections, particularly in countries like Mexico, should explore multi-party election models; these could be then compared to simpler models like the one presented in this study.

\section{Conclusions}
In this paper, we have explored nine election models in the context of the 2021 Mexican legislative election. Four of these models are variants of common approaches based on tweet volume and sentiment.  
We have examined a further four models which utilise geodata , and, we have also introduced a new positive-allegiance model. 
Most of the literature modelling elections relies on at most a few election models per study \cite{khan2021election}; here we implement several of the most commonly-used models and present direct comparisons of their predictions. 
Importantly, we have used bootstrapping statistics to include uncertainties in our vote-share estimates, a method that can be easily incorporated in other studies that rely on data from social media.
We find that models which  exclusively use geodata outperform all models that use the complete data except for the new positive-allegiance model. 
We find that the positive-allegiance model, which uses the full dataset, performs extremely well. 
The positive-allegiance model is intuitive, simple to implement, and it does not rely on geodata, which can be difficult to gather or might not be available at all. 
Our results suggest that it would be beneficial to use the positive-allegiance model in future election analyses. 
Additionally, we propose that analysis with geodata should be seriously considered in future analyses, as they might provide revealing insight to a problem at only a fraction of the data; for Twitter, today, that means an analysis that is $\sim$100 times faster (based on the subset of geodata out of the complete data).

The number of online users worldwide has been increasing since the creation of the internet \cite{ITU2021}.
In the last decade, the online population of adults in the United States has increased from $\sim$74\% to $\sim$93\% \cite{pew2021}.
In Mexico, the percentage of internet users aged 6 or more grew from 57\% in 2015 to 72\% in 2020, and is likely larger in 2022 \cite{ENDUTIH2020}.
While Mexico has not yet reached the online presence of, e.g., South Korea, the United Kingdom, and Sweden, and is more similar in this aspect to, e.g., Italy and Brazil, it has been steadily growing and becoming more connected \cite{ENDUTIH2020,ITU2021}.
Social media users have increased alongside internet coverage, and many studies have highlighted their role in politics \cite{BURNAP2016230,digrazia2013more,dimitrova2018social,bovet2018validation,radicioni2021analysing}.
Our study shows that positive-allegiance analysis and geo-analysis of Mexican elections via Twitter data is accurate, precise, and robust, and that there is a positive correlation between the population of Mexico, the number of internet users in Mexico, and the number of Twitter users discussing the 2021 Mexican legislative election.
This is particularly enlightening given that, while in Mexico the majority of the population is connected online, the percentage of online adults have not yet passed 90\%.
Our study finds a positive correlation between the online activity of the population of Twitter users in Mexico and real-life behaviour of Mexican citizens.
This suggests that countries similarly or more connected than Mexico 
have already transitioned into a period in time when online activity can be used to model and predict real-life behaviour.


\begin{backmatter}
    
\section*{Abbreviations}
\begin{itemize}
    \item API: Application Programming Interface
    \item NLTK: Natural Language Toolkit
    \item OSINT: Open-source Intelligence
\end{itemize}

\section*{Availability of data and material}
The tweet IDs and user IDs we used are available at \href{https://doi.org/10.5281/zenodo.7877001}{10.5281/zenodo.7877001} \cite{vigna_gomez_alejandro_2022_7877001}.
The scripts allowing to perform queries and data extraction for this manuscript, via the Twitter API for Academic Research, are available via GitHub at
\href{https://github.com/avigna/twitter-analysis}{avigna/twitter-analysis}.
The Python module used for the data extraction itself is available via GitHub at \href{https://github.com/DocNow/twarc}{DocNow/twarc} \cite{ed_summers_2022_6503180}.

\section*{Competing interests}
J.M., M.R., A.B., and I.M. are current employees of Metrics.
A.V-G. and P.K.R. report no competing interests.

\section*{Funding}
Not applicable.

\section*{Author's contributions}
A.V-G. and J.M. conceived the project. 
A.V-G., M.R. and I.M. performed the analysis and prepared the figures.
A.B. assisted with the data mining.
P.K.R. contributed to the analysis and interpretation of the data.
A.V-G. wrote most of the manuscript.
All authors contributed to the analysis, discussion, and writing of the paper.

\section*{Acknowledgements}
We thank Twitter API for Academic Research, under the project ``Representativeness of Social Behaviour Trends based on Twitter Data", for archival access to data.
We thank D. D'Orazio, I. Mandel, and I. Rivadeneyra for useful discussions. 
We thank J. Naiman, J. Vigna-G\'omez, and E. Wisbech for their comments on the manuscript.

\bibliographystyle{bmc-mathphys} 
\bibliography{sn-bibliography}





\section*{Figures}
\begin{figure}
\centering
\includegraphics[trim=0cm 0cm 0cm 0cm, clip,width=\textwidth]{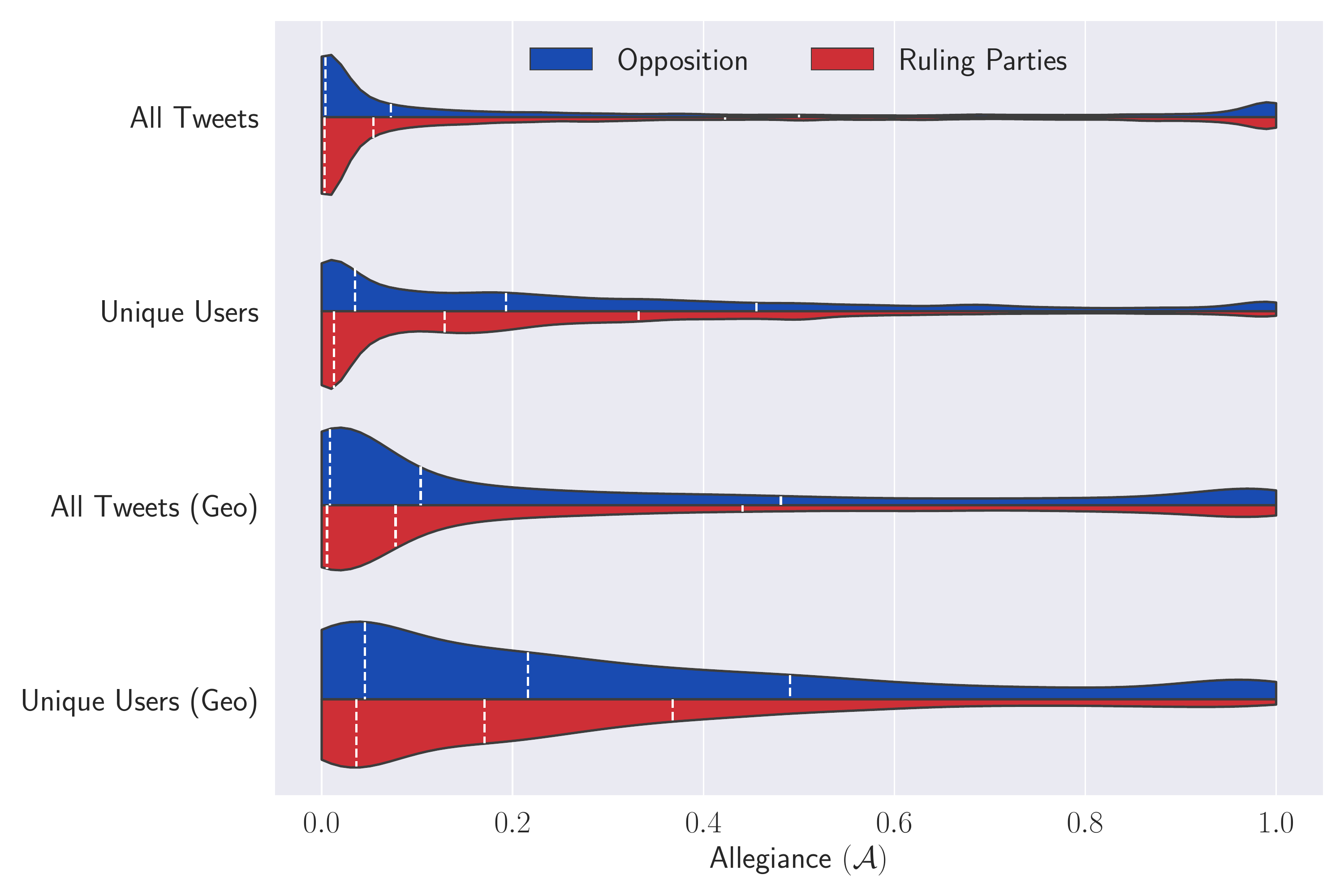}
\caption{
Violin plots comparing the determined allegiance ($\mathcal{A}$) of unique users to the ruling parties (red) and the opposition (blue), where $\mathcal{A}\approx 0$ means disapproval and implies a negative allegiance, while $\mathcal{A}\approx 1$ means approval and implies a positive allegiance.
The data is from May 2021, the month preceding Election Day.
For all the distributions, the bulk of the data is around $\mathcal{A}\approx 0$ and there are local maxima around $\mathcal{A}\approx 1$.
The number of tweets used to construct each plot decreases from top to bottom.
For the ruling parties(opposition), there is a total of 2.9(2.3) million messages, which encompass 304(201) thousand unique users, decreasing to 11(13) thousand messages with geodata posted by 2.4(2.7) thousand unique users.
We denote the quartiles in each distribution with white dashed lines.
}
\end{figure}

\begin{figure}
\centering
\includegraphics[trim=0cm 7cm 0cm 7cm, clip,width=\textwidth]{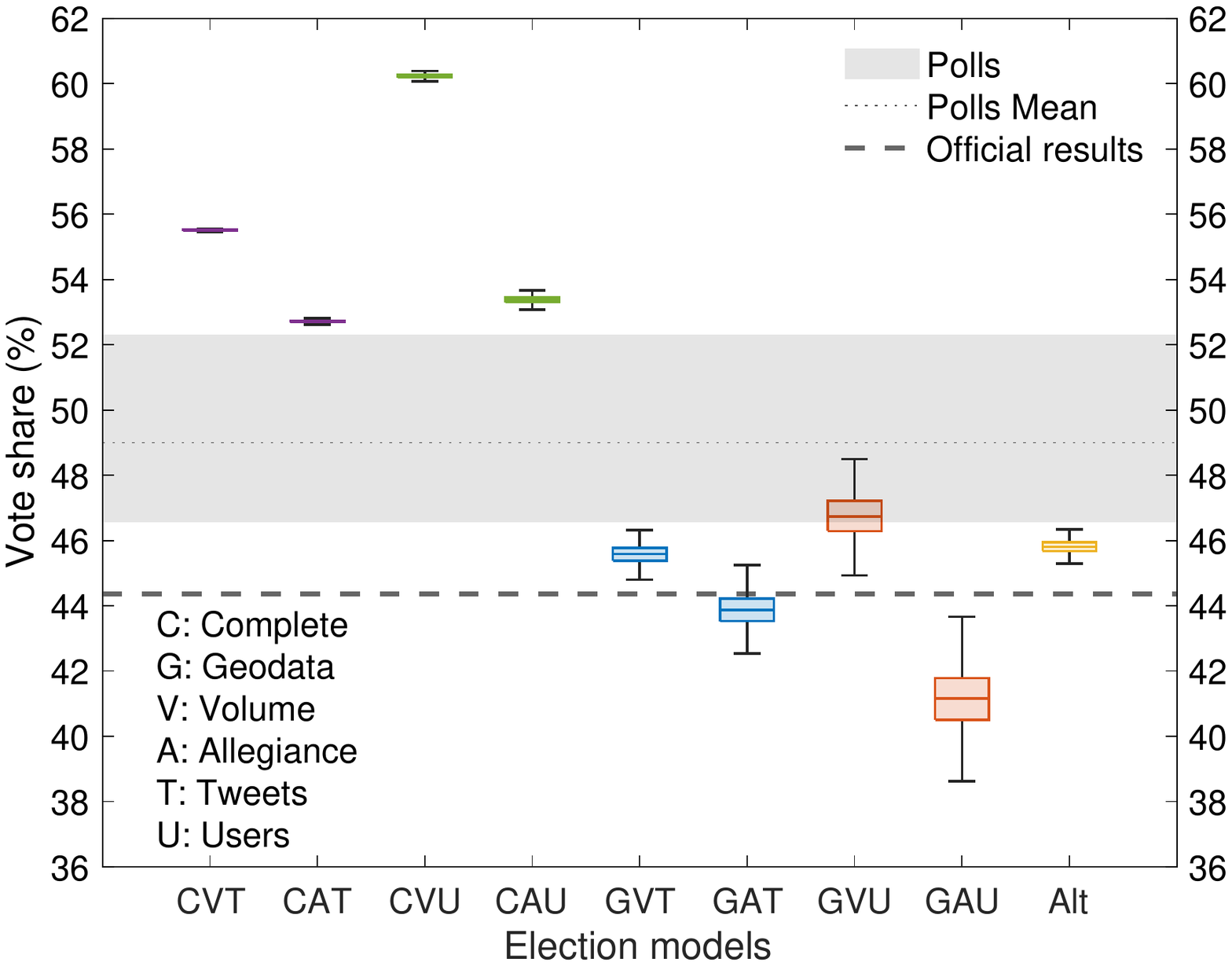}
\caption{
Election models and results of the 2021 Mexican legislative election.
The figure shows the vote-share for the ruling parties during May 2021 according to all models, as well as the official results (black thick dashed line) and results of polling aggregates (black thin dotted line) with reported uncertainties (grey shaded region).
The election models are shown as box plots defined by the median and the lower and upper quartiles.
Models using geodata (blue and red) outperform conventional polls.
Additionally, we present an alternative model (yellow) which considers positive-allegiance tweets exclusively (see Sect. 3.3 for more details).
Model nomenclature is defined with three letters as follows.
The first letter determines if the database considered is complete (C) or only accounts for tweets with geodata (G).
The second letter determines if the analysis is volumetric (V) or focuses on determined allegiances (A). 
The third letter determines if the analysis is performed in all tweets (T) or focuses on individual users (U).
}
\end{figure}

\begin{figure}
\centering
\includegraphics[trim=2cm 0cm 2cm 0cm, clip,width=\textwidth]{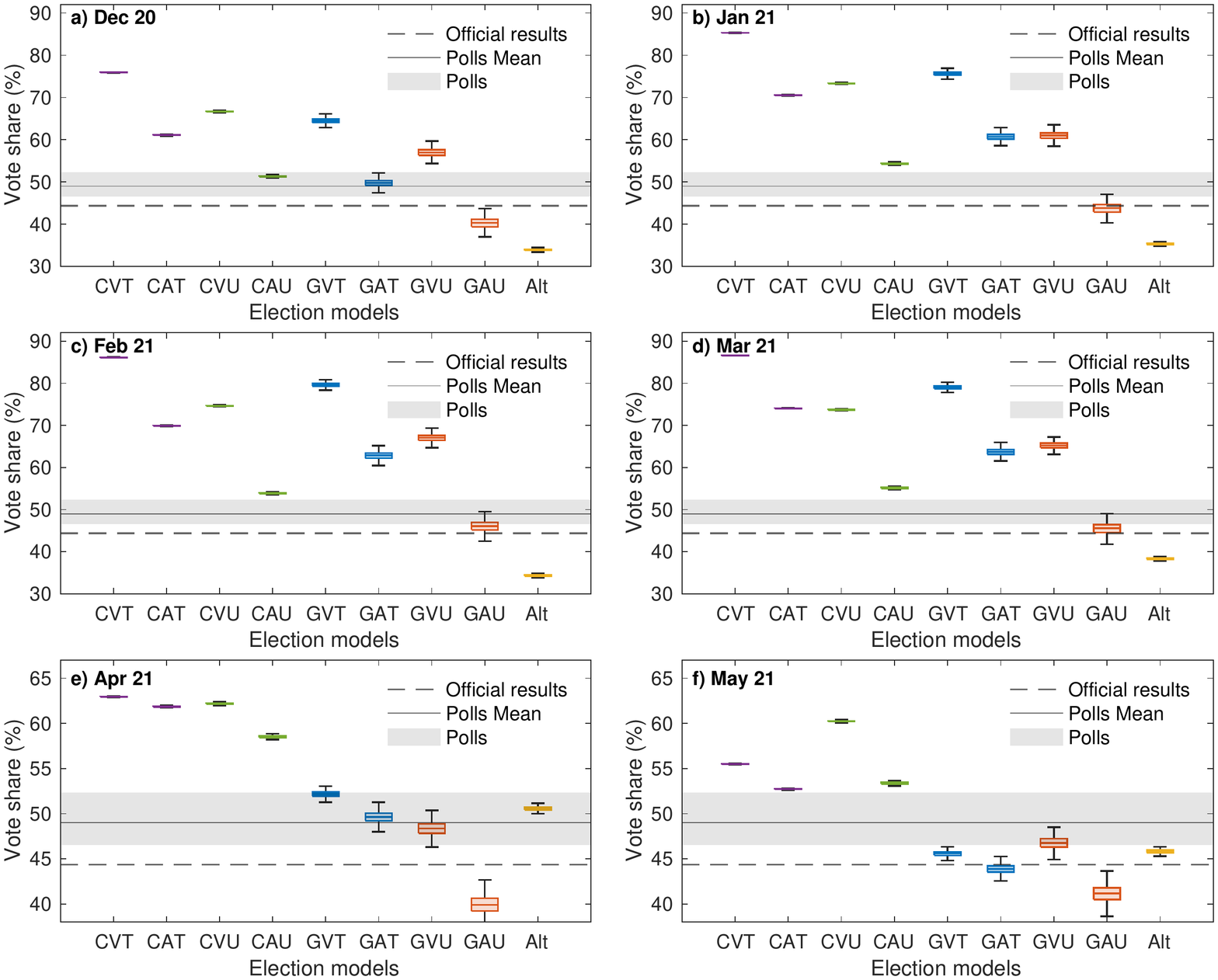}
\caption{
Vote-share monthly analysis of the 2021 Mexican legislative election.
Color and model nomenclature is the same as presented in Figure 2.
}
\end{figure}

\begin{figure}
\centering
\includegraphics[trim=0cm 7cm 0cm 7cm, clip,width=\textwidth]{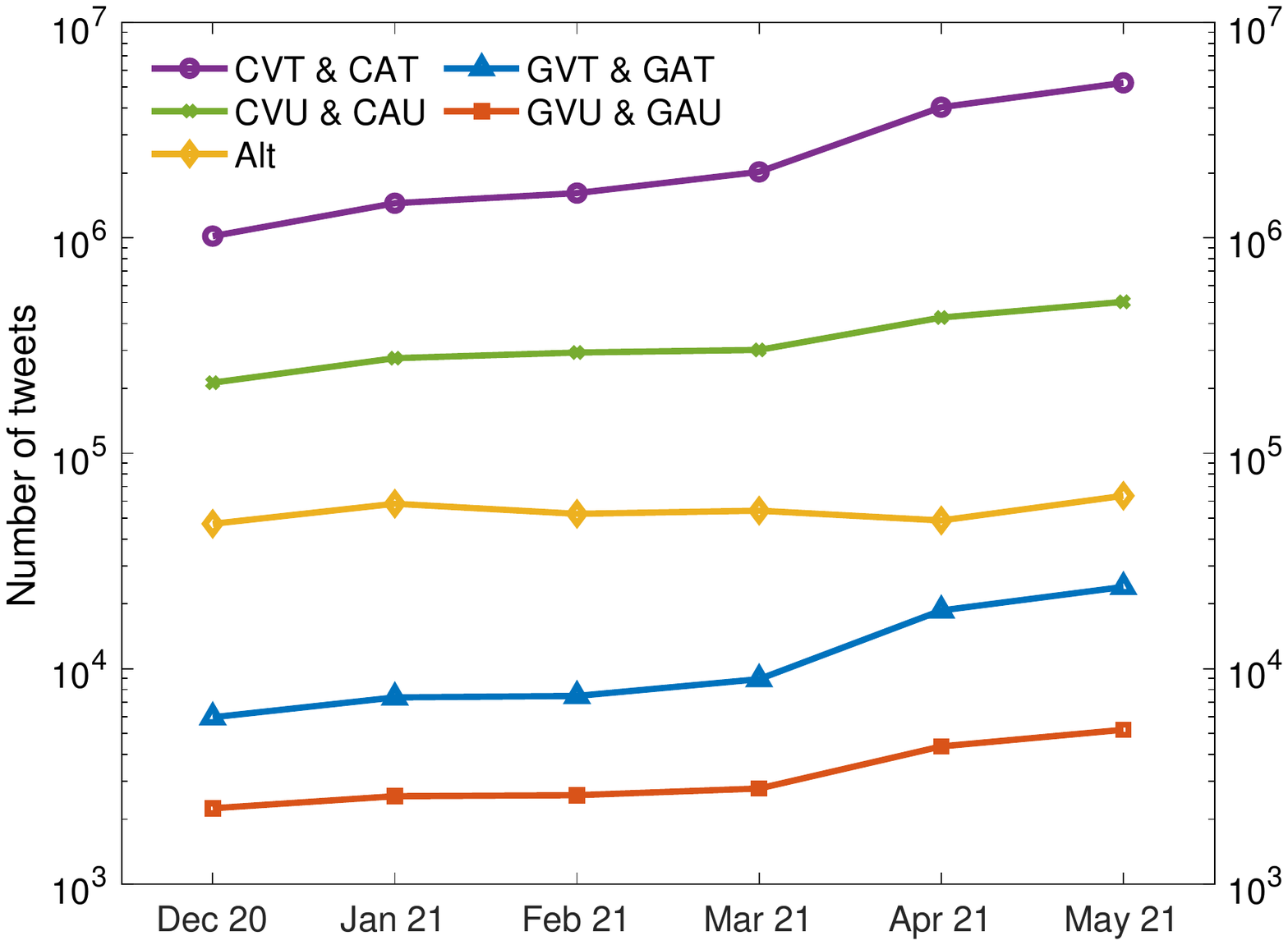}
\caption{
Monthly tweets related to the 2021 Mexican legislative election.
Model nomenclature is the same as presented in Figure 2.
}
\end{figure}

\begin{figure}
\centering
\includegraphics[trim=0cm 7cm 0cm 7cm, clip,width=0.65\columnwidth]{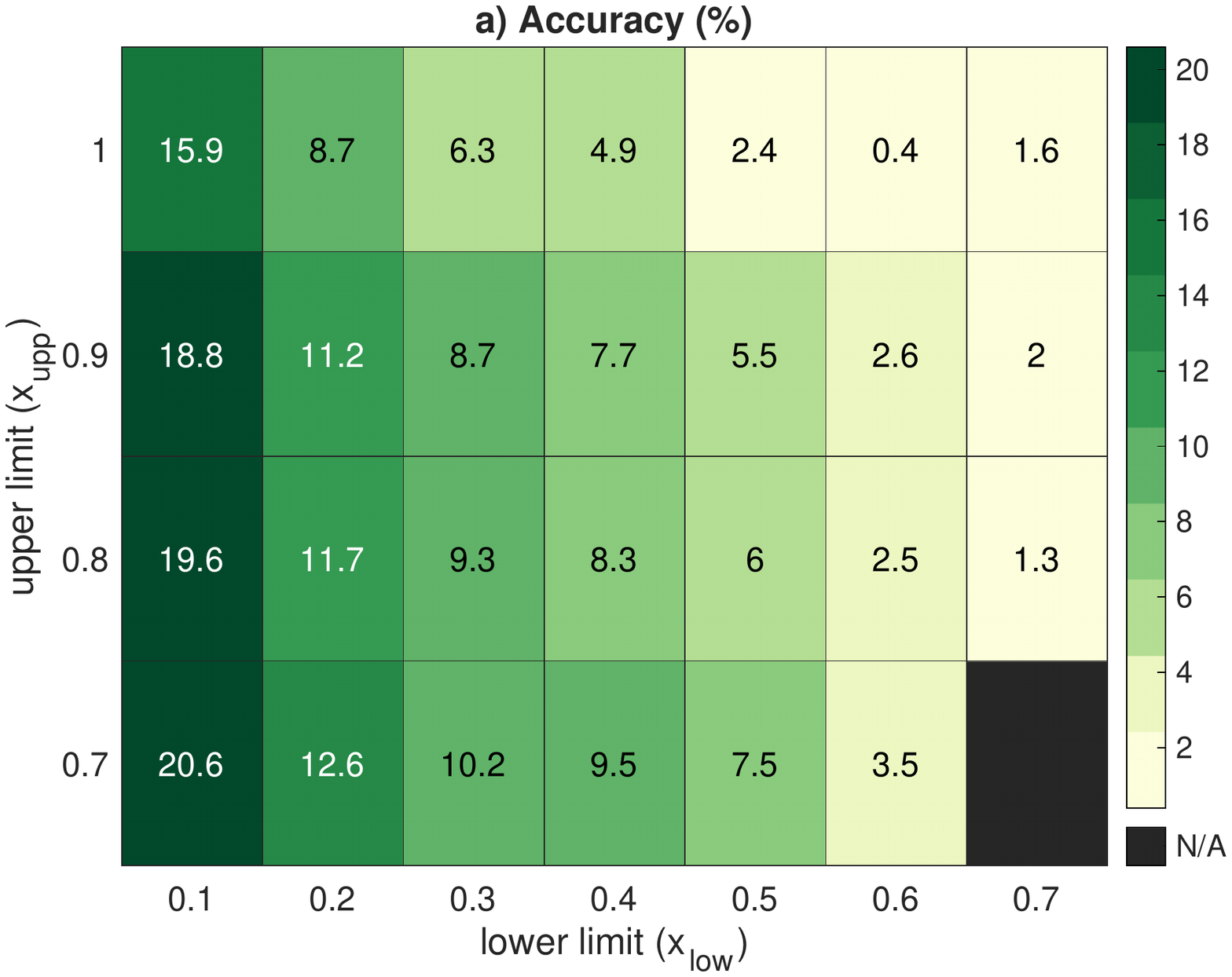}
\includegraphics[trim=0cm 7cm 0cm 7cm, clip,width=0.65\columnwidth]{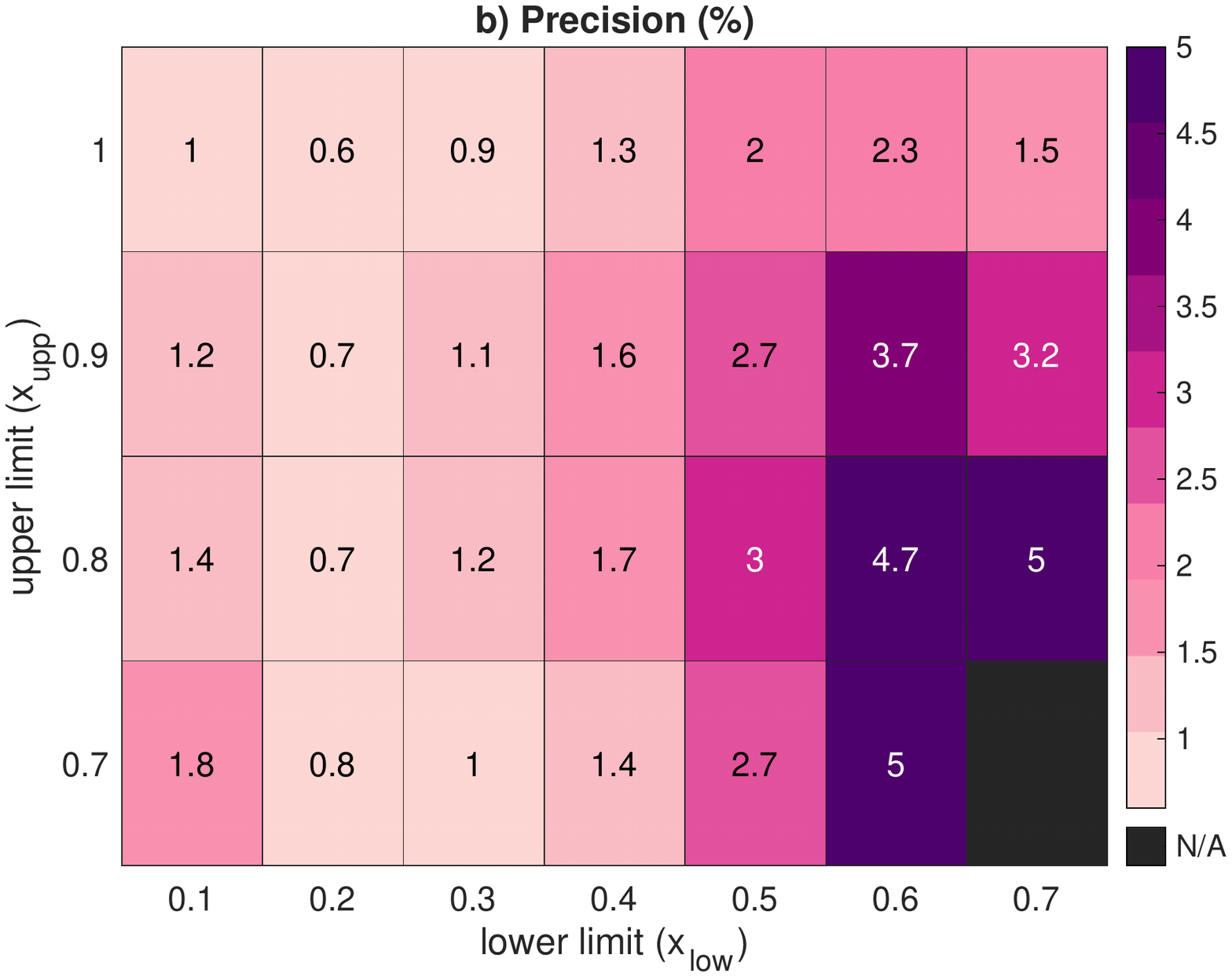}
\includegraphics[trim=0cm 7cm 0cm 7cm, clip,width=0.65\columnwidth]{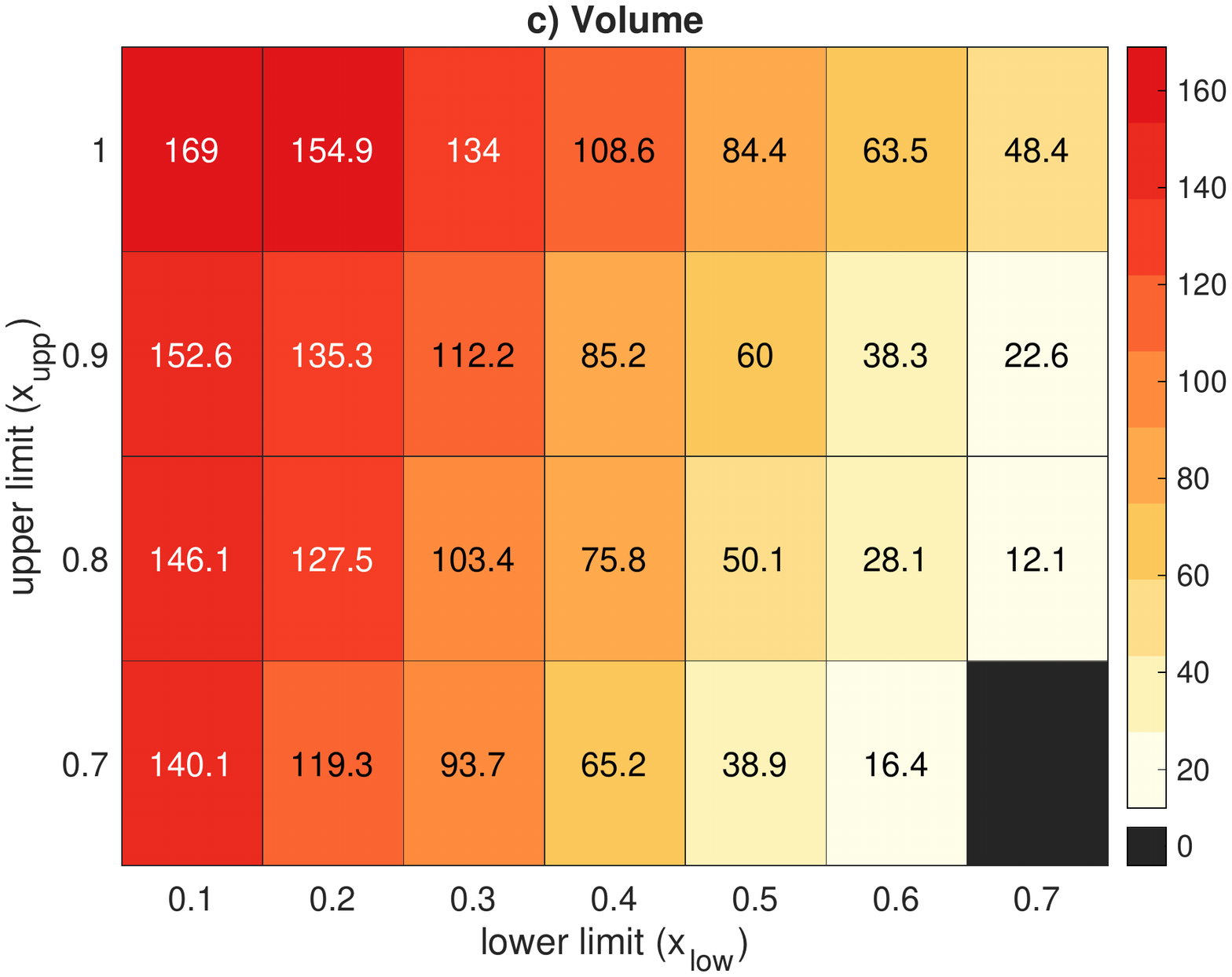}
\caption{
Analysis of the alternative election model for May 2021.
In this model, we determine political preference based on positive allegiance, that we define with respect to a lower limit ($x_{\rm{low}}$, horizontal axis) and an upper limit ($x_{\rm{upp}}$, vertical axis).
Our default sub-model assumes $\{ x_{\rm{low}}, x_{\rm{upp}} \}=\{0.6, 1.0\}$ and therefore $\mathcal{A} \geq 0.6$.
Panel \textbf{a} shows accuracy as the percentage difference between our model and the selection results.
Panel \textbf{b} shows precision as the maximum bootstrapping uncertainties, in percentage points. 
Panel \textbf{c} shows volume as the number of thousands of unique users.
}
\end{figure}

\begin{figure}
\centering
\includegraphics[trim=0cm 7cm 0cm 7cm, clip,width=1.0\textwidth]{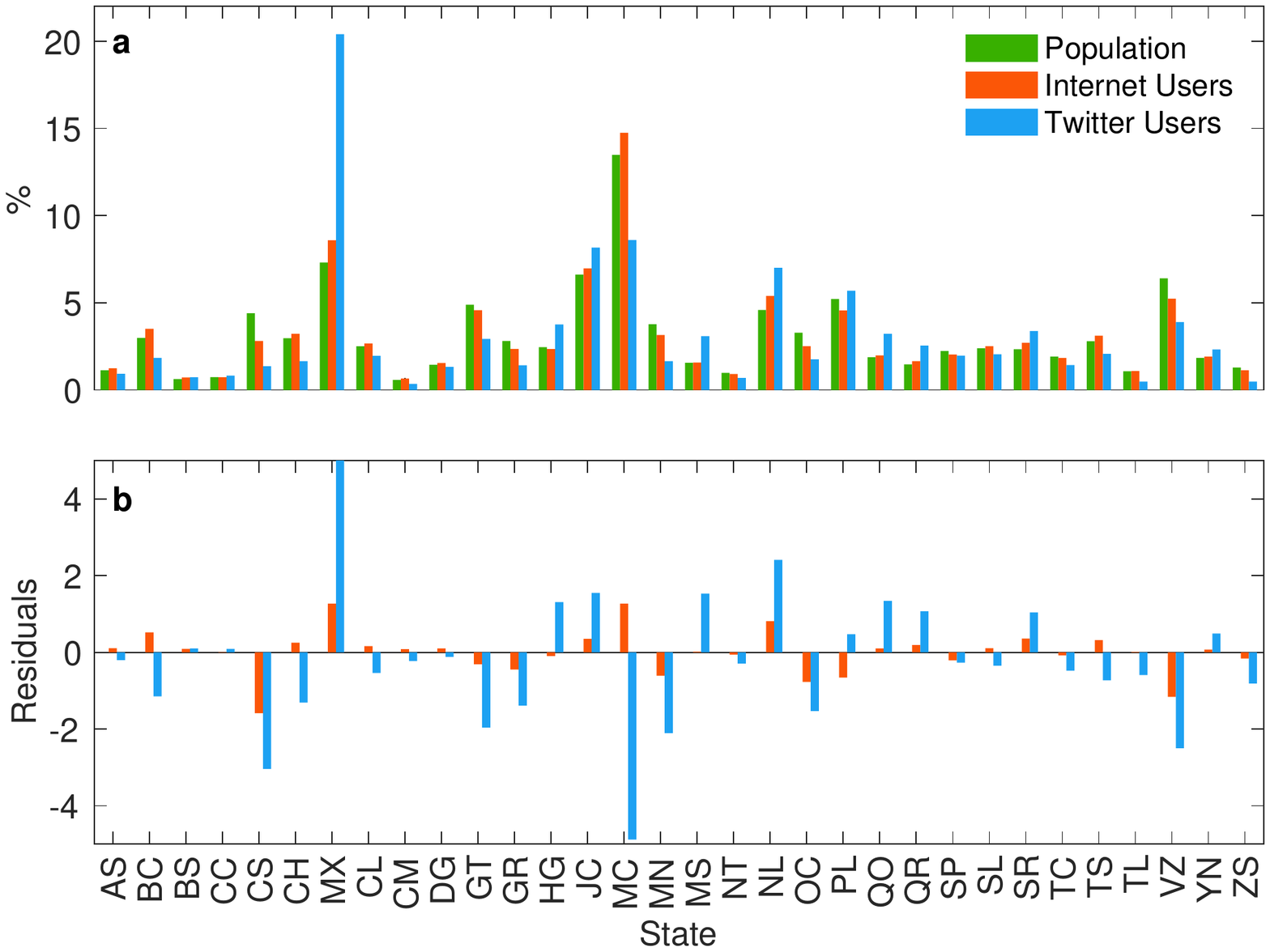}
\caption{
Quantitative geographical analysis using official census results from 2020 and the subset ($\approx$0.5\%) of all of our Twitter data that contains geodata.
Panel \textbf{a} shows the percentage of the population of Mexico (green), internet users in Mexico (red) and Twitter users in our data (blue) for the 32 states of Mexico.
Mexico City (MX) is clearly over-represented and an outlier from the Twitter sample.
The Pearson’s correlation coefficient (r) between the percentage of the population of Mexico and percentage of internet users in Mexico is $r=0.98$, and decreases to $r=0.67$ when comparing to our Twitter data. 
This value increases to $r=0.96$ when we combine the data from MX, Hidalgo (HG) and the State of Mexico (MC), which encompasses the conurbation around Mexico City known as Greater Mexico City.
The data are correlated.
Panel \textbf{b} shows the residuals of the population of Mexico with respect to internet and Twitter users.
This is done in a range where Mexico City is not visible.
For the bulk of the data, the percentage of internet users is representative of the population within $\lesssim 1.6\%$, while Twitter data is representative within $\lesssim 3.0\%$.
}
\end{figure}

\begin{figure}
\centering
\includegraphics[trim=0cm 7cm 0cm 7cm, clip,width=\textwidth]{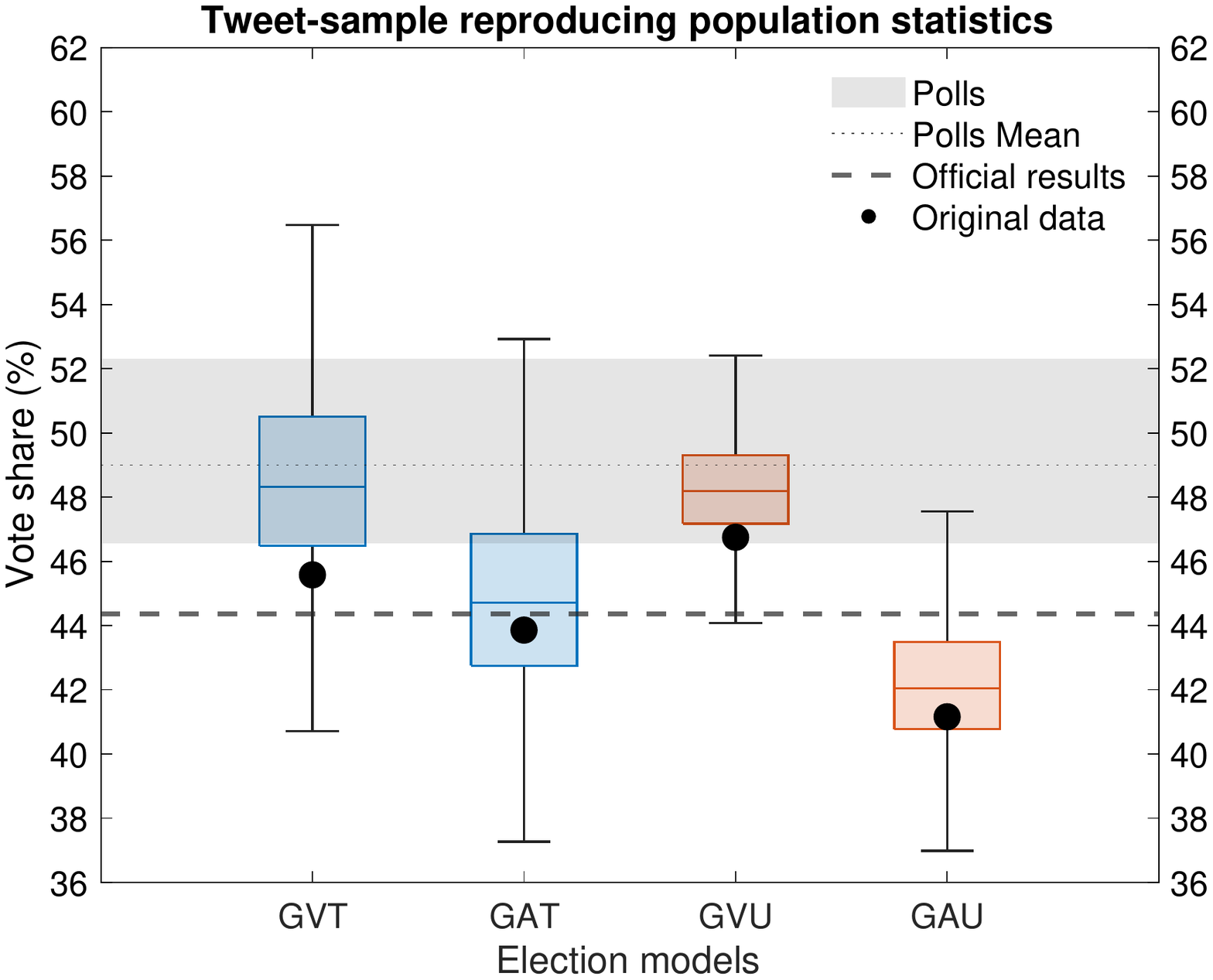}
\caption{
Analysis of the results of the 2021 Mexican legislative election.
Here we present only the models which include geodata.
For each model, we present the results as solid black circles.
The boxplots correspond to alternative realizations of the original distribution of each model, where 1000 users are drawn to match the population distribution of Mexico (Figure 6).
Results are in very good agreement, highlighting the correlation of the Twitter geodata with respect to the population.
Color and model nomenclature are the same as presented in Figure 2.
}
\end{figure}


\section*{Tables}
\begin{table}
\caption{
Summary of the testing and training accuracy of our allegiance determination model. For each party/coalition, we show the total number of tweets, total number of messages of the training set, as well as the number of them that have a negative (n) or a positive (p) connotation. We also provide the $F_1$ and ROC AUC score.
}
\centering
\begin{tabular}{ |l|c|cc|cc|c| } 
\hline\hline
\multirow{2}{*}{Party} & \# of tweets & \multicolumn{2}{|c|}{\# of messages} & \multicolumn{2}{|c|}{$F_1$ score} & \multirow{2}{*}{AUC}  \\
& & n & p & n & p & \\
\hline
MORENA + PT & 10,357,147 & 405 & 405 & 0.61 & 0.69 & 0.79\\
PVEM & 239,126 & 65 & 65 & 0.82 & 0.67 & 0.85 \\
PAN & 1,474,845 & 200 & 200 & 0.47 & 0.68  & 0.61 \\
PRI & 1,248,188 & 235 & 235 & 0.84 & 0.82 & 0.91 \\
PRD & 1,374,029 & 302 & 302 & 0.82 & 0.79 & 0.85 \\
MC & 418,066 & 231 & 231 & 0.72 & 0.72 & 0.84 \\
PES + FxM + RSP & 262,472 & 285 & 285 & 0.57 & 0.52 & 0.66 \\
\hline\hline
\end{tabular}
\label{app:table}
\end{table}




\end{backmatter}
\end{document}